# Backward Pulling Force from a Forward Propagating Beam

Jun Chen,[1] Jack Ng,[2*] Zhifang Lin,[3] and C. T. Chan[2]

[1]State Key Laboratory of Surface Physics and Department of Physics,
Fudan University, Shanghai 200433, China

[2]Department of Physics, The Hong Kong University of Science and Technology, Kowloon, Hong Kong

[3]Key Laboratory of Micro and Nano Photonic Structures (Ministry of Education),
Fudan University, Shanghai, China

*E-mail: jack@ust.hk

*A photon carries a momentum of $\hbar k$, so one may anticipate light to "push" on any object standing in its path via the scattering force.[1,2,3,4] In the absence of intensity gradient, using a light beam to pull a particle backwards is counter intuitive. Here, we show that it is possible to realize a backward scattering force which pulls a particle all the way towards the source without an equilibrium point. The underlining physics is the maximization of forward scattering via interference of the radiation multipoles. We show explicitly that the necessary condition to realize a negative (pulling) optical force is the simultaneous excitation of multipoles in the particle and if the projection of the total photon momentum along the propagation direction is small (as in some propagation invariant beams), attractive optical force is possible. This possibility adds "pulling" as an additional degree of freedom to optical micromanipulation.*

It is well known that light can push an object forward. A focused light beam can also "trap" particles as small particles will migrate to the intensity maxima[5,6] as in the case of optical tweezers when the gradient force due to intensity inhomogeneity overcomes the forward scattering force of the photons. These "push" and "trap" functionalities are the basis of modern optical micromanipulation.[7,8,9] Using a beam of light to pull a particle towards the source via backward scattering force (BSF) is counter-intuitive and indeed, it can be shown straightforwardly that BSF is not possible for a plane wave. The Poynting vector can be "reversed" in some Bessel beams and it has been suggested that the reversed Poynting vector can lead to attractive optical forces.[10] However, a careful consideration (see supplementary information) would show that a reversed Poynting vector is neither a sufficient nor a necessary condition for realizing BSF. In the following, we shall establish the conditions to achieve BSF for certain optical beams. For simplicity, we shall consider a propagation invariant beam (PIB), such as a Bessel beam (BB),[1,2,11,12,13,14,15,16,17,18,19] which has vanishing gradient force along the propagation (*z*) axis. The Fourier decomposition of a PIB consists of plane wave components whose *k*-vectors form a cone that make an angle $\theta_0$ with the *z*-axis.[20,21] Thus the *z* component momentum of the incident photons is $\hbar k \cos\theta_0$. In a scattering event, the momentum is scattered elastically in another direction, and the momentum of the recoiled photons is, on average, $\hbar k \langle\cos\theta\rangle$, where $\langle\cos\theta\rangle \leq 1$ is the asymmetry parameter.[22] It is derived in the supplementary information that

$$F = W_{sca} c^{-1} \left( \cos\theta_0 - \langle\cos\theta\rangle \right), \quad (1)$$

where $W_{sca}$ is the rate at which photon energy is scattered and *c* is the speed of light. The first term in equation (1) corresponds to the beam momentum removed by the particle. For a plane wave, $\cos\theta_0 = 1$, so BSF is impossible. The *z*-component of the momentum of incident photons decreases as $\theta_0$ increases, and *F*, as given by equation (1), may not be positive definite. However, it is by no means obvious that *F* can be negative at any value of $\theta_0$, since the second term may also vanish with $\theta_0$. The value of $\langle\cos\theta\rangle$, being a function of the beam profile as well as the particle's optical properties, describes the "weighted average" direction of the scattered radiation. Only if the radiation is emitted predominantly in the forward direction, a large backward recoil force may give rise to a BSF.

To analytically demonstrate that BSF is possible, let us derive the multipole expansion (see supplementary information)[23] of the time-averaged optical force up to electric quadrupole order:

$$\vec{F} = \vec{F}_{incident} + \vec{F}_{interference}, \quad (2)$$

where

$$\begin{aligned}\vec{F}_{incident} &= \vec{F}_{\vec{p}} + \vec{F}_{\vec{m}} + \vec{F}_{\tilde{Q}_e} + ..., \\ \vec{F}_{interference} &= \vec{F}_{\vec{p}\vec{m}} + \vec{F}_{\tilde{Q}_e \vec{p}} + ...,\end{aligned} \quad (3)$$

and

$$\vec{F}_{\vec{p}} = \tfrac{1}{2}\operatorname{Re}\{\nabla \vec{E}^* \cdot \vec{p}\}, \qquad \vec{F}_{\vec{m}} = \tfrac{1}{2}\operatorname{Re}\{\nabla \vec{B}^* \cdot \vec{m}\},$$
$$\vec{F}_{\vec{Q}_e} = \tfrac{1}{4}\operatorname{Re}\{\nabla\nabla \vec{E}^* : \vec{\vec{Q}}_e\}, \qquad (4)$$
$$\vec{F}_{\vec{p}\vec{m}} = -\tfrac{k^4}{12\pi\varepsilon_0 c}\operatorname{Re}\{\vec{p}\times\vec{m}^*\}, \quad \vec{F}_{\vec{Q}_e\vec{p}} = -\tfrac{k^5}{40\pi\varepsilon_0}\operatorname{Im}\{\vec{\vec{Q}}_e \cdot \vec{p}^*\},$$

Here $k$ is the wavenumber, $\vec{E}$ and $\vec{B}$ are the incident fields, $\vec{p}$ and $\vec{\vec{Q}}_e$ are the electric dipole and quadrupole moments, and $\vec{m}$ is the magnetic dipole moment. Every term in $\vec{F}_{\text{incident}}$ are product of a multipole and the incident field or field gradient. They represent forces exerted directly by the incident wave. On the other hand, every term in $\vec{F}_{\text{interference}}$ are product of two multipole moments. They represent forces induced by the interference of the multipoles' radiation fields (see supplementary information). Now consider the $z$ component forces, since the gradient force vanishes, only the scattering force acts on the particle. It can be shown that (see supplementary information)

$$\left(\vec{F}_{\text{incident}}\right)_z = W_{sca} c^{-1} \cos\theta_0,$$
$$\left(\vec{F}_{\text{interference}}\right)_z = -W_{sca} c^{-1} \langle\cos\theta\rangle. \qquad (5)$$

Equation (5) indicates that $\left(\vec{F}_{\text{incident}}\right)_z$ is positive definite, accordingly, any negative force must come from $\left(\vec{F}_{\text{interference}}\right)_z$, which is induced by the interference of the induced multipoles. For appropriate phases in the multipole moments, the interference may cause the particle to emit more light forward to produce a larger $\langle\cos\theta\rangle$, resulting in a large recoil force that overcomes $\vec{F}_{\text{incident}}$. To see this explicitly and analytically, consider an incident BB with incident $E$ field,[16]

$$\vec{E} = E_0 e^{ik_z z + im\phi} \left\{ \begin{array}{l} \left[im\alpha J_m(k_\perp\rho)/\rho + ik_z\beta k_\perp J'_m(k_\perp\rho)/k\right]\hat{\rho} \\ -\left[k_z m\beta J_m(k_\perp\rho)/k\rho + \alpha k_\perp J'_m(k_\perp\rho)\right]\hat{\phi} + \left[\beta k_\perp^2 J_m(k_\perp\rho)/k\right]\hat{z} \end{array} \right\}, \qquad (6)$$

where $(\rho,\phi,z)$ are the cylindrical coordinates, $k_\perp = k\sin\theta_0$, $\alpha = \eta_{\text{TE}} ik/k_\perp^2$, and $\beta = \eta_{\text{TM}} k e^{i\eta}/k_\perp^2$. Here, $\eta$ is the relative phase between TM and TE waves, $J_m$ is a Bessel function of order $m$, and $J'_m$ is the Bessel function's derivative with respect to its argument. For the first order BB characterize by $m=1$, $\eta_{\text{TE}} = \sqrt{2}$, $\eta_{\text{TM}} = 3$, and $\eta = \pi/2$, when equation (6) is substituted into equation (2) and retaining leading terms up to $(kr_s)^8$ (with $r_s$ the particle radius), one arrives at

$$\left(\vec{F}_{incident}\right)_z \simeq 4\pi\varepsilon_0 \cos\theta_0 E_0^2 \frac{(\varepsilon_r-1)^2}{k^2(\varepsilon_r+2)^2}$$
$$\times \left[\frac{1}{3}(kr_s)^6 + \frac{2(\varepsilon_r-2)}{5(\varepsilon_r+2)}(kr_s)^8\right] + O(kr_s)^9, \quad (7)$$

$$\left(\vec{F}_{interference}\right)_z \simeq -\frac{2\pi\varepsilon_0 E_0^2(\varepsilon_r-1)^2}{45k^2(\varepsilon_r+2)(2\varepsilon_r+3)}$$
$$\times \left[3\sqrt{2}\varepsilon_r + (6+11\varepsilon_r)\cos\theta_0\right](kr_s)^8 + O(kr_s)^9. \quad (8)$$

Here, $\mu_r = 1$ is the permeability, $\varepsilon_r$ is the permittivity. The multipole moments needed in equations (2)-(4) are given by $\vec{p} = \alpha_e \vec{E}$, $\vec{m} = \alpha_m \vec{B}$, $\vec{\vec{Q}}_e = (\gamma_e/2)(\nabla \vec{E} + \nabla \vec{E}^T)$, $\alpha_e = i6\pi\varepsilon_0 a_1/k^3$, $\alpha_m = i6\pi b_1/k^3\mu_0$, $\gamma_e = i40\pi\varepsilon_0 a_2/k^5$, and ($a_1$, $a_2$, $b_2$) are the Mie coefficients[22]. Note that $\left(\vec{F}_{incident}\right)_z > 0$ as the leading term in equation (7), which $\sim(kr_s)^6$, is positive. Moreover $\left(\vec{F}_{interference}\right)_z < 0$ for some range of $\varepsilon_r$ (for example, $\varepsilon_r > 0$). Consequently, the sign of the force $F_z$ is determined by the competition between $\left(\vec{F}_{incident}\right)_z$ and $\left(\vec{F}_{interference}\right)_z$. When $\theta_0$ approaches $\pi/2$, $\left(\vec{F}_{incident}\right)_z \sim \cos\theta_0$ while $\left(\vec{F}_{interference}\right)_z$ approaches a constant. $\left(\vec{F}_{interference}\right)_z$ will therefore become dominant, leading to $F_z<0$ for certain ranges of $\varepsilon_r$. This demonstrates analytically that under appropriate conditions a BB, or a PIB, can act as an "optical tractor beam".

The optical force changes sign when $\theta_0 > \theta_c$ for some critical angle, $\theta_c$, which can be obtained by putting the description of a first order BB into equations (3), (4) and (6).[24] In Figure 1, $\theta_c$ is plotted as a function of $\varepsilon_r$ and $kr_s$, with $\mu_r = 1$, $\eta_{TE} = \sqrt{2}$, $\eta_{TM} = 3$, and $\eta = \pi/2$. No physical solution of $\theta_c$ can be found for a large part of the region with $\varepsilon_r < 0$ (the grey region), meaning that no BSF is possible. In this $\varepsilon_r < 0$ region, the strong reflection is not favorable to maintaining a BSF. A small part of the region with $\varepsilon_r < 0$ has physical $\theta_c$ solutions. This can be attributed to plasmon resonances at which $\vec{F}_{interference} = \vec{F}_{\vec{p}\vec{m}} + \vec{F}_{\vec{Q}_e\vec{p}}$ overcomes $\vec{F}_{incident} = \vec{F}_{\vec{p}} + \vec{F}_{\vec{m}} + \vec{F}_{\vec{Q}_e} \approx \vec{F}_{\vec{p}}$. ($\vec{F}_{\vec{m}}$ and $\vec{F}_{\vec{Q}_e}$ are small compared with the other terms that are linear in $\bar{p}$.) For $\varepsilon_r > 0$, $\theta_c$ always exists, as in the analytical equations (7) and (8).

Consider a first order BB with some specific $\theta_0$ acting on a spherical particle with dielectric properties specified by $\varepsilon_r$ and $\mu_r$. The calculated forces are shown in Fig. 2, where the white regions indicate positive forces and the colored regions indicate a BSF. Figure 2a presents the situation with $\theta_0 = 64°$ and $kr_s = \pi/5$. A BSF can be observed when $\varepsilon_r \approx \mu_r$, because impedance matching with the air minimizes backward reflection, which facilitates a BSF. Figure 2a-b are not perfectly symmetric about the diagonal line where $\varepsilon_r = \mu_r$ because perfect symmetry requires that when we exchange $\varepsilon_r$ and $\mu_r$, we will also need to exchange the *E*- and *H*-fields. At $\theta_0 = 64°$, a BSF is observed only in the first and third quadrants. In the second and forth quadrants the wave cannot propagate within the particle as the refractive index is imaginary, resulting in an enhancement of reflection that reduces $\langle \cos\theta \rangle$, which is not favorable to BSF.

Figure 2b shows that the BSF allowed region shrinks when the sphere radius increases from $kr_s = \pi/5$ to 2. Large particle size is unfavorable to a BSF, because as the particle size approaches the geometric optics limit, interference effects may be neglected. The particle then cannot maximize the forward scattering by interference and thus no BSF. In Fig. 2c, the BSF region shrinks when $\theta_0$ decreases from 64° to 58° due to the larger projection of the incident photons' momentum onto the *z*-axis. In Fig. 2c, a BSF still survives in some regions of the phase space despite of the relatively small $\theta_0$. In these regions the amplitude of the electric dipole, magnetic dipole, and electric quadrupole are simultaneously large and they can thus interfere to produce strong forward scattering. This also explains why these regions are narrow, because the three modes can interfere constructively only for some rather specific material parameters.

For the Rayleigh particles shown in Fig. 2, there is no BSF at $\theta_0 = 64°$ when $\mu_r \approx 1$. This is because, in essence, such particle possesses only the electric dipole mode, so $\bar{F}_{\text{interference}} \simeq 0$ and its $\theta_c$ at $\mu_r \approx 1$ approaches $\pi/2$. This is not the case for Rayleigh particles with $\mu_r \neq 1$ or for larger dielectric particles in the Mie regime, as these particles also possess higher order multipole moments. At optical frequencies, $\mu_r = 1$ for the great majority of materials. Accordingly, except for large $\theta_0$, BSF is almost exclusively limited to particles in the Mie regime, excluding Rayleigh particles and large particles. An exception is particles made from metamaterials[25] that have a magnetic response at optical frequencies.

We now show that BSF can be realized in simple dielectric particles. The possibility of a BSF in "ordinary" particles such as polystyrene beads is shown numerically in Fig. 3 where we plot the computed optical forces acting on polystyrene beads illuminated by a BB as a function of particle radius. A BSF is possible at multiple radii when the particle has no absorption ($\text{Im}\{\varepsilon_r\} = 0$, red dashed line). If some losses are added ($\text{Im}\{\varepsilon_r\} = 0.01$, line connecting blue circles), the force remains more or less the same when the particle is small. Larger particles, however, absorb more photons, and each photon absorbed (and not re-emitted) imparts forward momentum, eventually ruining the BSF for large particles.

The emergence of BSF can be visualized in Fig. 4, where the angular dependence of the normalized scattered irradiance (defined as the energy scattered per unit time into a unit solid angle normalized by $W_{sca}$, the rate at which the photon energy is scattered) is given as a polar plot for two particle sizes, one subject to pulling (red line) and one subject to pushing (blue line). The green dashed line gives the direction $\theta_0$ as defined in equation (1). If a particle scatters the incident photons predominately in directions with $\theta < \theta_0$ (i.e. to the right side of the green dashed line), the net momentum gained by the particle is negative and the particle would be "attracted" by the beam. This is the case for a polystyrene bead with $r_s = 2.03\mu m$, as shown by the red line. We also show for comparison the normalized scattered irradiance for a polystyrene bead with $r_s = 2.19\mu m$ (blue line) which is pushed by the beam.

Finally, if one is to use, for example, a BB to pull a particle backward using BSF, the particle has to be stably confined in the transverse directions. The transverse stability of the particle can be determined through a linear stability analysis.[26,27] In Fig. 3, the black curve highlights the regimes in which a BSF and stable transverse trapping are observed simultaneously, and thus these particle sizes are capable of long distance stable backward transportation. Since the direction of the force is size dependent, BSF may be employed for particle sorting.

In summary, this analysis established that light can indeed pull a particle, and the conditions under which an "attractive" scattering force can be observed. BSF arises from the interference of multipoles excited in particles interacting with the beam. This analysis has shown both analytically and numerically that a BSF can also be observed in certain particles that have both electric and magnetic responses or in dielectric particles in the Mie regime. In water, a objective lens with N.A.=1.3 can support $\theta_0$ up to ~77.8°, which is more than sufficient to produce BSF. A large $\theta_0$ in a PIB is always favorable for observing backward scattering, but the underlining physics suggests that a BSF is possible for some specific particles for any $\theta_0 > 0$. Note that BSFs are not limited to PIBs. A BSF can in general arise in beams composed of near glancing plane wave components, and this may open up new avenues for optical micromanipulation (see supplementary information), of which typical examples include transporting a particle backward over a long distance and particle sorting.

**Methodology**

We calculate the time-averaged optical force that acts on a spherical particle via a surface integral of the time averaged Maxwell stress tensor over the sphere's surface. The electromagnetic fields needed in the Maxwell stress tensor are computed by the rigorous and accurate Generalized Lorentz-Mie theory, where the field quantities are expanded in a series of vector spherical wave functions. This series expansion is truncated at some angular momentum $L_{max}$, which is chosen such that further increase in the number of terms in the series does not alter the optical force. This generalized Lorentz-Mie theory and Maxwell stress tensor formalism[27,28,29] can be considered as "ab initio", in the sense that no approximation is needed (up to numerical truncation). In all numerical calculations, the intensity at the beam center is normalized to $1\,\mathrm{mW/\mu m^2}$.

**Acknowledgement**

We acknowledge support from Hong Kong's Research Grants Council (GRF grant# 600308), from the National Natural Science Foundation of China (NSFC) (10774028) and the Chinese Ministry of Education (B06011). Computational resources were supported by the Shun Hing Education and Charity Fund and the HPCCC at Hong Kong Baptist University.

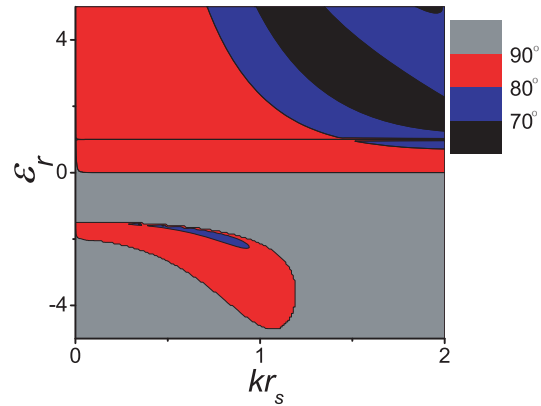

**Figure 1 | A contour plot for the minimum beam angle $\theta_c$ to observe backward scattering force**. The grey region represents conditions where the force cannot be negative at any $\theta_0$. For $kr_s=0$ or $\varepsilon_r =1$, effectively there is no particle, so $\theta_c$ is meaningless.

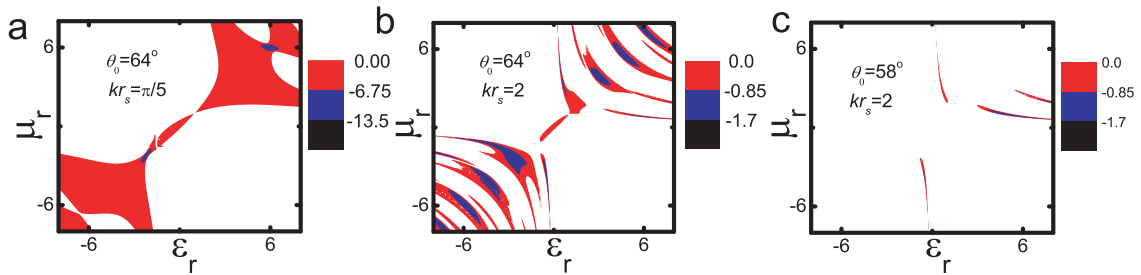

**Figure 2 | Existence of BSF for particles made up of different material.** The calculated optical force in arbitrary units as a function of relative permittivity $\varepsilon_r$ and relative permeability $\mu_r$. White regions indicate positive optical force and colored regions indicate the parameter space where the optical force is negative (i.e. BSF). In the

calculation, $m=1$, $\lambda=1064nm$, $\eta_{TE}=\sqrt{2}$, $\eta_{TM}=3$, and $\eta=\pi/2$. **a**, $\theta_0=64^0$ and $kr_s=\pi/5$. **b**, $\theta_0=64^0$ and $kr_s=2$. **c**, $\theta_0=58^0$ and $kr_s=2$.

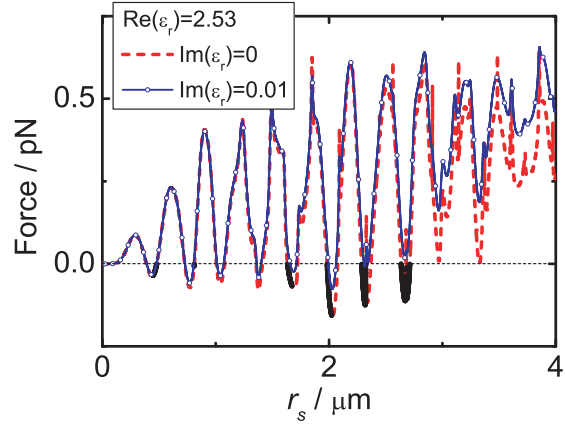

**Figure 3 | Existence of backward scattering force on a polystyrene sphere illuminated by a Bessel beam.** The optical force as a function of $r_s$ for a spherical polystyrene particle ($\text{Re}\{\varepsilon_r\}=2.53$) illuminated by a BB with $m=0$, $\lambda=1064nm$, $\eta_{TE}=0$, $\eta_{TM}=1$, $\eta=0$, and $\theta_0=78.5^0$. The dashed red line and the line with blue circles denote $\text{Im}\{\varepsilon_r\}=0$ and $\text{Im}\{\varepsilon_r\}=0.01$ respectively. The black curve represents regions with both a BSF and stable transverse trapping.

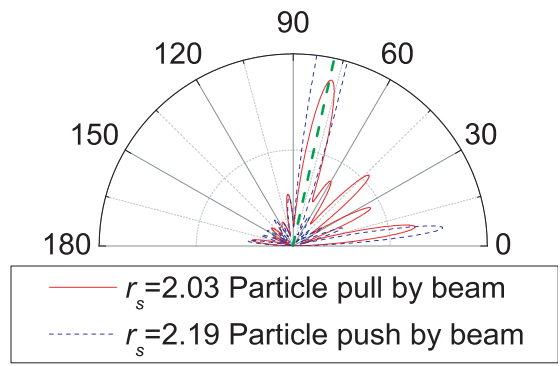

**Figure 4 | Polar plot for normalized scattered irradiance showing the angular dependence of scattered light:** The incident beam is the same as that of Fig. 3. The angular dependence of the scattered irradiance is plotted as a function of $\theta$ as the scattering is independent of $\phi$. The green dotted line marks the direction of $\theta_0$ defined in

equation (1), which indicate the incident beam's momentum projected onto the beam axis. The red line is the normalized scattered irradiance for a particle with $r_s = 2.03\mu m$. The scattered light is predominately scattered in forward directions with $\theta < \theta_0$ and the beam attracts the particle. The blue line is for the particle ($r_s = 2.19\mu m$). In this case, the light beam pushes.